\documentclass{elsart}
\usepackage{epsfig}

\begin{document}
\begin{frontmatter}

\title{Effects of mass media on opinion spreading in the Sznajd sociophysics model}
\author{Nuno Crokidakis\thanksref{nuno}}
\address{
Instituto de F\'{\i}sica - Universidade Federal Fluminense \\
Av. Litor\^anea, s/n \\
24210-340 \hspace{5mm} Niter\'oi - Rio de Janeiro \hspace{5mm} Brazil}

\thanks[nuno]{nuno@if.uff.br}

\maketitle

\begin{abstract}

In this work we consider the influence of mass media in the dynamics of the two-dimensional Sznajd model. This influence acts as an external field, and it is introduced in the model by means of a probability $p$ of the agents to follow the media opinion. We performed Monte Carlo simulations on square lattices with different sizes, and our numerical results suggest a change on the critical behavior of the model, with the absence of the usual phase transition for $p>\sim 0.18$. Another effect of the probability $p$ is to decrease the average relaxation times $\tau$, that are log-normally distributed, as in the standard model. In addition, the $\tau$ values depend on the lattice size $L$ in a power-law form, $\tau\sim L^{\alpha}$, where the power-law exponent depends on the probability $p$.

\end{abstract}
\end{frontmatter}


\section{Introduction}

\quad Social dynamics have been studied through statistical physics techniques in the last twenty years. Among the studied problems, we can cite models of cultural \cite{axelrod}, language \cite{baronchelli} and opinion dynamics \cite{holley,galam,sznajd} (for a recent review, see \cite{loreto_rmp}). These kinds of models are interesting to physicists because they present order-disorder transitions, scaling and universality, among other typical features of physical systems \cite{loreto_rmp}.

One of the most studied models of opinion dynamics in the last years is the Sznajd model \cite{sznajd,sznajd_review}. The original Sznajd model \cite{sznajd} consists of a chain of sites with periodic boundary conditions where each site (individual opinion) could have two possible states (opinions) represented in the model by Ising spins (``yes'' or ``no''). A pair of parallel spins on sites $i$ and $i+1$ forces its two neighbors, $i-1$ and $i+2$, to have the same orientation (opinion), while for an antiparallel pair $(i,i+1)$, the left neighbor ($i-1$) takes the opinion of the spin $i+1$ and the right neighbor ($i+2$) takes the opinion of the spin $i$. In this first formulation of the model two types of steady states are always reached: complete consensus (ferromagnetic state) or stalemate (anti-ferromagnetic state), in which every site has an opinion that is different from the opinion of its neighbors. However, the transient displays an interesting behavior, as pointed by Stauffer et al. \cite{adriano}. 

A more interesting situation was studied in \cite{adriano}, where the model was defined on a $L\times L$ square lattice. The authors in \cite{adriano} considered not a pair of neighbors, but a $2\times 2$ plaquette with four neighbors. Considering that each plaquette convince its eight neighbors if all four center spins are parallel, and that the initial density of up spins is $d=0.5$, the authors found that the system reaches the fixed points with all up or all down spins with equal probability. For $d<0.5$ ($>0.5$) the system goes to a ferromagnetic state with all spins down (up) in all samples, which characterizes a phase transition at $d=0.5$ in the limit of large $L$. This phase transition separates two distinct states of the system: for $d<0.5$ the system never reaches full-consensus states with all spins up, whereas for $d>0.5$ the consensus is always reached. Other formulations of the two-dimensional model, considering for example memory \cite{sabatelli}, reputation \cite{meu_pla}, diffusion of agents \cite{adriano2} or a random dilution of the lattice \cite{auto}, also present similar phase transitions.

The utility of the Sznajd model goes beyond the basic description of a community \cite{sznajd_review}. It was applied, for example, to politics. In 1999, Costa-Filho et. al. \cite{raimundo} showed that distribution of votes per candidate for the 1998 elections in Brazil follow a power law with exponent $\sim -1.0$. Based on the Sznajd model, two models with more than two opinions were proposed, considering the dynamics on square and cubic regular lattices, and also in complex Barab\'asi networks \cite{barabasi}. The authors successfully reproduced the distribution of the number of candidates according the number of votes they received in Brazilian elections \cite{sznajd_elections1,sznajd_elections2,sznajd_elections3}. The Sznajd model was also applied to marketing, where it was considered advertising and a competition of two different products, and to finance, where the authors found a good agreement between some characteristics of the price trajectories (like returns, for example) and the simulations \cite{sznajd_review}. It was also claimed that some other characteristic of Sznajd-like models may also be present in real social systems, like the power-law relationship between the time need to reach the fixed point (the complete consensus), called the relaxation time, and the system size \cite{adriano2}.

However, we can observe that the above-mentioned works did not take into account the effects of an external field in this phase transition. In a real community, external effects may be, for example, the mass-media influence. It is well-known that, in real life, the mass media (television, radio, ...) has a great influence in the population, and people tend to keep or change their opinions on any question according to that influence. These effects were considered in some social models, specially in the Axelrod model of cultural diversity \cite{mazzitello1,mazzitello2,rodriguez,pabjan,gonzalez1,gonzalez2}, and interesting results were found. Thus, in this work we consider the effects of mass media in the opinion dynamics of the Sznajd sociophysics model. This influence is introduced in the model by means of a probability $p$ of the agents to follow the media opinion. We performed Monte Carlo (MC) simulations for different lattice sizes, and our results suggest that the system undergoes a phase transition, as well as in the original Sznajd model, for values of the probability $p<\sim 0.18$, and that the relaxation times are log-normally distributed. In addition, the average relaxation times $\tau$ depend on the lattice size $L$ in a power-law form $\tau\sim L^{\alpha}$, where $\alpha$ is a function of $p$.

This work is organized as follows. In Section 2 we present the microscopic rules that define the model. The numerical results are discussed in Section 3, and our conclusions are presented in Section 4.


\section{Model}

\begin{figure}[t]
\begin{center}
\includegraphics[width=0.2\textwidth,angle=0]{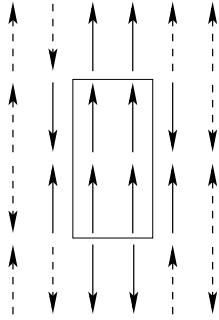}
\end{center}
\caption{Schematic representation of a plaquette (spins inside the rectangle) and its eight neighbors (full-line spins). Notice that the other agents (dotted-line spins) do not participate in the dynamics.}
\label{Fig0}
\end{figure}

We have considered the Sznajd model defined on a square lattice with linear size $L$ and periodic boundary conditions. The lattice sites were numbered by one index $i$, $i=1,2,...,N$, where $N=L^{2}$ is the total number of agents in the population. We assign an Ising variable to each site, $s_{i}=\pm 1$, representing the two possible opinions of each agent. At each time step, the following $3$ microscopic rules control our model (see the schematic representation of Fig. \ref{Fig0}):
\begin{enumerate}
\item We \textit{randomly} choose a 2 $\times$ 2 plaquette of four neighbors.
\item If all four center spins are parallel, the eight nearest neighbors are persuaded to follow the plaquette orientation.
\item If not all four center spins are parallel, we consider the influence of mass media: each one of the eight neighbors follows, \textit{independently} of the others, the media opinion with probability $p$.
\end{enumerate}

Notice that we update the agents' states in a random sequential order (asynchronous updating). After the above three steps, we count one MC step in our model. In addition, we will consider that the mass media is favorable to the opinion $+1$, i.e., in the above-mentioned rule (3), each neighbor (independently of the others) will change his opinion to $+1$ with probability $p$ if he was not persuaded by the plaquette agents (i.e., if not all four plaquettes' spins are parallel). In other words, we will take into account that the influence of a group of agents is stronger than the influence of media. In fact, we can imagine that in the real world people tend to be more influenced by friends, relatives and colleagues (represented in the model by the plaquettes' agents), among others, in comparison with the influence of the media (represented in the model by the probability $p$). Notice that for $p=0$ we recover the standard Sznajd model defined on the square lattice \cite{adriano}.

As pointed by Stauffer in \cite{stauffer_review}, we can imagine that each agent in the Sznajd model carries an opinion, that can either be $+1$ or $-1$ and that represents one of two possible opinions on any question. For example, in an election, the opinion $+1$ may represent the intention of an individual to choose a certain candidate A, whereas the opinion $-1$ may represent the intention to choose another candidate B. Thus, the inclusion of the probability $p$ in the model will take into account the influence of TV and/or advertisement, for example, to provoke a change in the choice (i.e., the candidate) of that individual.


\section{Monte Carlo Simulation}

\begin{figure}[t]
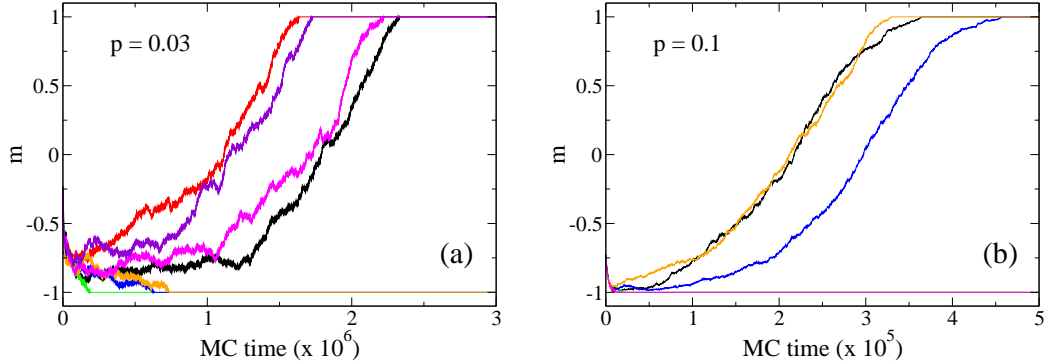

\begin{center}
\includegraphics[width=0.47\textwidth,angle=0]{figure2a.eps}
\hspace{0.4cm}
\includegraphics[width=0.47\textwidth,angle=0]{figure2b.eps}
\end{center}
\caption{(Color online) Time evolution of the magnetization per site $m$, Eq. (\ref{eq1}), for $L=73$ and different samples. The parameters are $p=0.03$, $d=0.38$ (a) and $p=0.1$, $d=0.13$ (b). Observe that in both figures the consensus with all spins pointing up is achieved in some samples, even for $d<<0.5$, that is the critical density for the standard model \cite{adriano}.}
\label{Fig1}
\end{figure}

Following the previous works on the Sznajd model, we can start studying the time evolution of the magnetization per site,
\begin{equation}\label{eq1}
m=\frac{1}{N}\sum_{i=1}^{N}s_{i}~,
\end{equation}
\noindent
where $N=L^{2}$ is the total number of agents in the population and $s_{i}=\pm 1$. In the standard Sznajd model defined on the square lattice \cite{adriano}, if we consider an initial density of up spins $d=0.5$ and that a $2\times 2$ plaquette with all spins parallel persuades its eight neighbors, the system reaches the fixed points with all up or all down spins with equal probability. For $d<0.5$ ($>0.5$) the system goes to a ferromagnetic state with all spins down (up) in all samples, which characterizes a phase transition at a critical value $d_{c}=0.5$ in the limit of large $L$. We can observe that for $p>0$ this result changes (see Fig. \ref{Fig1}). In Fig. \ref{Fig1}(a) we exhibit the simulation data for $p=0.03$ and initial density of up spins $d=0.38$. In this case, we observe that in some realizations of the dynamics the system reaches consensus with all spins up, even for $d<0.5$. In the case of a greater value of $p$, for example $p=0.1$, the difference between our model and the standard one \cite{adriano} is more pronounced, and a strong effect of the media influence is observed: even for a very small value of the initial density of up spins ($d=0.13$) we observe that the full-consensus state with $m=1$ is obtained in some samples [see Fig. \ref{Fig1}(b)]. Notice that as in the original 2D Sznajd model \cite{adriano}, the system always reaches consensus states with all up or all down spins, i.e., there are no states with magnetization per spin $|m|<1$. However, the presence of an external field (mass media) changes the critical behavior of the 2D Sznajd model. In addition, if a phase transition also occurs in our case, the position of the transition points $d_{c}$ will depend on $p$.

\begin{figure}[t]
\begin{center}
\includegraphics[width=0.7\textwidth,angle=0]{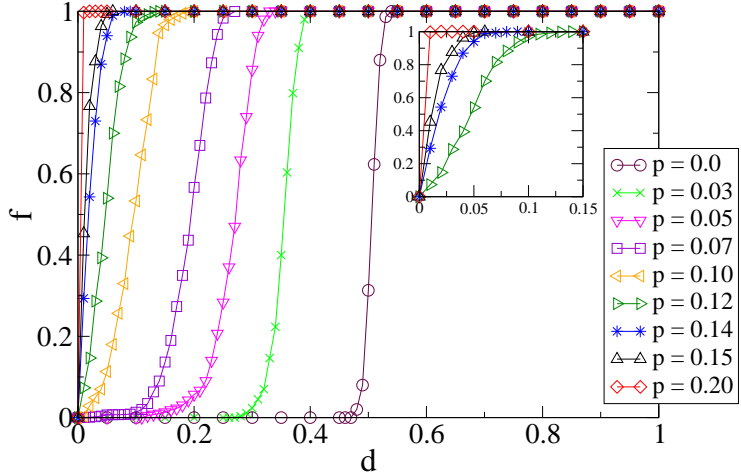}
\end{center}
\caption{(Color online) Fraction $f$ of samples which show all spins up when the initial density of up spins $d$ is varied in the range $0.0\leq d\leq 1.0$ for $L=121$ and typical values of the probability $p$. In the inset we show results for values in the range $0.12\leq p \leq 0.2$. Observe that the transition does not occur for $p=0.2$, i.e., we have $f(d=0)=0$ and $f(d>0)=1$. We have considered $300$ samples for each point in the graphics.}
\label{Fig2}
\end{figure}

The next step is to analyze if the phase transition observed in the standard 2D Sznajd model also occurs in our version of the model. In other words, given a certain value of the probability $p$, we want to determine if there is a critical value $d_{c}$ above which the system always reaches full-consensus states with $m=1$. For this purpose, we have simulated the system for different lattice sizes $L$ and typical values of the probability $p$, and we have measured the fraction of samples which show all spins up when the initial density of up spins $d$ is varied in the range $0.0 \leq d \leq 1.0$. In other words, this quantity $f$ gives us the probability that the population reaches consensus for a given value of $d$. The results for the largest size considered, $L=121$, are exhibited in Fig. \ref{Fig2}, where we have considered $300$ samples for each point. We can observe the transition from $f=0$ to $f=1$ at different values of $d$, for a given value of $p$. However, for $p=0.2$ the transition does not occur, i.e., we have that $f=1$ even for very small values of the initial density of up spins like $d=0.01$ (see the inset of Fig. \ref{Fig2}). The exact threshold value $p_{{\rm thres}}$ above which the system does not present the phase transition is difficult to determine numerically. Nonetheless, our simulations suggest that the phase transition only occurs in our formulation of the Sznajd model for $p<\sim 0.18$. 

\begin{figure}[t]
\begin{center}
\vspace{0.7cm}
\includegraphics[width=0.6\textwidth,angle=0]{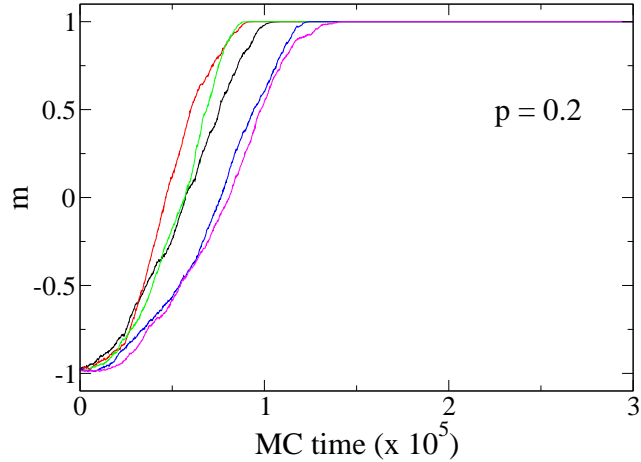}
\end{center}
\caption{(Color online) Time evolution of the magnetization per site $m$ for $L=73$, $p=0.2$, $d=0.01$ and different samples. Observe that for this value of the probability $p$, even for $d=1\%$ the full consensus with all spins pointing up is achieved in all samples, suggesting the absence of a phase transition.}
\label{FigX}
\end{figure}

To see this picture in more details, we performed simulations for $L=121$ and $d=0.01$, and the results obtained were $f=0.97$ for $p=0.18$, $f\sim 0.98$ for $p=0.19$ and $f=1.0$ for $p=0.2$, whereas for $p=0.16$ and $p=0.15$ we obtained $f=0.69$ and $f=0.46$, respectively. We can obtain the same conclusion monitoring the time evolution of the magnetization per site $m$. For example, in Fig. \ref{FigX} we exhibit the magnetization $m$ as a function of the MC time for $d=0.01$, $p=0.2$ and different samples. Observe that even for a very small value of the initial density of up spins the full consensus is achieved in all samples, suggesting the absence of a phase transition (similar results were found for $0.18$ and $0.19$). In other words, the mass media (i.e., the probability $p$) acts in the system in a way that it changes the opinion of the majority of the agents in the population, and the full-consensus steady states with $m=1$ are always achieved for $p>\sim 0.18$. As above-discussed, we are interested in the transition between a phase where the full-consensus states with all spins pointing up never occur and a phase where this consensus always occurs. In this sense, the main effect of the mass media is to eliminate the phase transition for $p>\sim 0.18$.

\begin{figure}[t]
\begin{center}
\vspace{0.6cm}
\includegraphics[width=0.6\textwidth,angle=0]{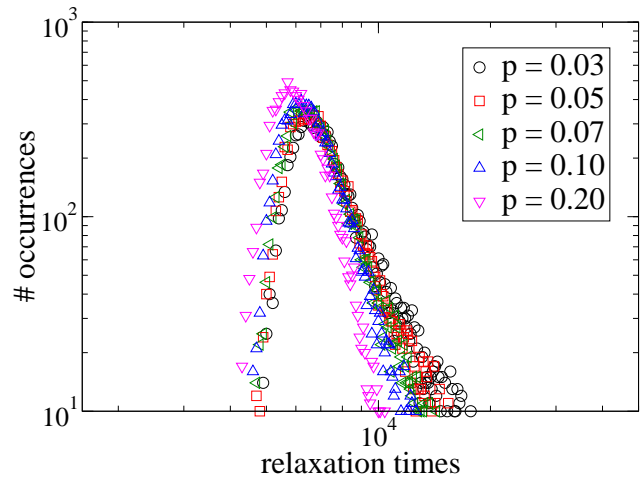}
\end{center}
\caption{(Color online) Log-log plot of the histogram of relaxation times for $L=73$, $d=0.5$ and some values of the probability $p$, obtained from $10^{4}$ samples. The distribution is compatible with a log-normal one for all values of $p$, which corresponds to the observed parabola in the log-log plot.}
\label{Fig3}
\end{figure}

\begin{figure}[t]
\begin{center}
\includegraphics[width=0.6\textwidth,angle=0]{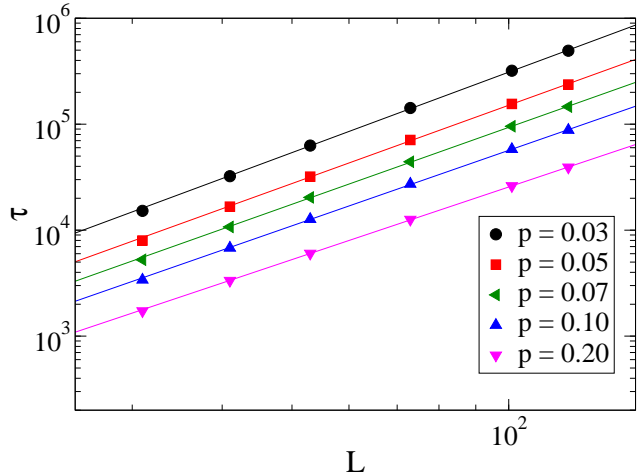}
\end{center}
\caption{(Color online) Average relaxation times $\tau$, over $10^{4}$ samples, versus lattice size $L$ in the log-log scale for $d=0.5$ and some values of the probability $p$. The straight lines are fittings that present different slopes, i.e., $\tau$ depends on $L$ in the power-law form $\tau\sim L^{\alpha}$, where $\alpha$ is a function of $p$.}
\label{Fig4}
\end{figure}

An usual quantity measured in MC simulations of Sznajd models is the relaxation time, i.e., the time needed to find all the agents at the end having the same opinion. The distribution of the number of sweeps through the lattice, averaged over $10^{4}$ samples, needed to reach the fixed point is shown in Fig. \ref{Fig3}. We can see that the relaxation times distribution is compatible with a log-normal one for all values of $p$, which corresponds to a parabola in the log-log plot of Fig. \ref{Fig3}. In addition, the effect of the probability $p$ is to decrease the relaxation times. In fact, the peaks of the distributions in Fig. \ref{Fig3} move to lower values of the relaxation times when we increase $p$. Notice that relaxation times log-normally distributed were observed in other versions of the Sznajd model \cite{adriano,meu_pla,adriano2,schneider}.

Based on the relaxation times distribution, we measured the average relaxation times $\tau$ (also over $10^{4}$ samples) as a function of the lattice size $L$. The results are exhibited in Fig. \ref{Fig4}, in the log-log scale. As in previous works on the Sznajd model \cite{meu_pla,adriano2}, we have found a power-law relation between these quantities,
\begin{equation}\label{eq2}
\tau\sim L^{\alpha}
\end{equation}
\noindent
but with different exponents $\alpha$, depending on the value of the probability $p$. Observe in Fig. \ref{Fig4} that the average relaxation times are smaller for increasing values of the probability $p$, which is a consequence of the effects of $p$ on the distribution of the relaxation times. In fact, we can observe that the peaks of the distributions in Fig. \ref{Fig3} move to lower values of the relaxation times when we increase $p$, as discussed above.

\begin{figure}[t]
\begin{center}
\includegraphics[width=0.6\textwidth,angle=0]{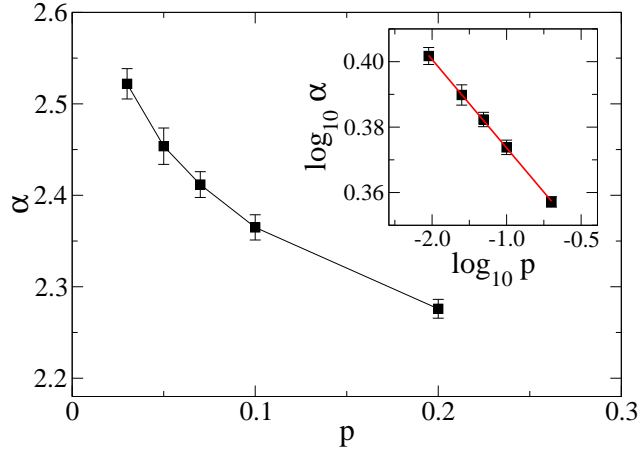}
\end{center}
\caption{(Color online) Exponent $\alpha$ of the average relaxation time, Eq. (\ref{eq2}), as a function of the probability $p$. In the inset we show the same data plotted in the log-log scale. Fitting data, we obtained $\alpha\sim p^{-0.05}$.}
\label{Fig5}
\end{figure}

In Fig. \ref{Fig5} we exhibit the exponent $\alpha$ as a function of $p$, and we can observe that $\alpha$ decreases for increasing values of $p$. It is also shown (see the inset of Fig. \ref{Fig5}) the same data in the log-log scale. Fitting data, we obtained another power-law behavior, 
\begin{equation}\label{eq3}
\alpha\sim p^{-0.05} ~.
\end{equation} 

It is important to say that the results presented in this work are robust against changes in some details of the model. For example, if we consider plaquettes with two agents (instead of four) that try to convince their six neighbors (instead of eight), we also observe the same results.

\section{Conclusions}

In this work, we have studied a modified version of the two-dimensional Sznajd consensus model. In particular, we have introduced the effects of mass media on the opinion dynamics of the model. This influence acts as an external field, and it is introduced in the model by means of a probability $p$ of the agents to follow, independently of the others, the media opinion. The standard Sznajd model is recovered for $p=0$. We performed Monte Carlo simulations of the model defined on square lattices, and our results are summarized in the following.

We have analyzed in detail the time evolution of the magnetization per spin for different values of the probability $p$. The simulations suggest that the system undergoes a phase transition only for values of $p<\sim 0.18$. This result was confirmed by measurements of the fraction of samples which show all spins up when the initial density of up spins $d$ is varied, that is the usual order parameter for the Sznajd model. Thus, the main effect of the probability $p$, that acts in the model as a mass-media (external) influence, is to eliminate the phase transition. This behavior is similar to the effect of an external uniform magnetic field $H$ on the 2D Ising model. The difference is that for the Ising model any value $H>0$ provokes the destruction of the ferromagnetic-paramagnetic phase transition \cite{stanley_book}, whereas in our case it is necessary an "external field" $p\sim 0.18$ to eliminate the phase transition of the 2D Sznajd model.

Observe that the inclusion of simple ingredients modify drastically the behavior of the Sznajd model. In fact, as discussed in the Introduction, the one-dimensional Sznajd model does not present a phase transition. However, the inclusion of a temperature provokes the appearance of a transition even in the 1D case \cite{la_lama}. Considering the two-dimensional model, the simplest formulation exhibits a phase transition \cite{adriano}, whereas the inclusion of an external field destroys this transition, as discussed in this work.

In order to measure the relaxation times of the model, we performed Monte Carlo simulations on square lattices with linear sizes up to $L=121$ and typically $10^{4}$ samples. As in the standard model \cite{adriano}, we found that the relaxation times are log-normally distributed. In addition, the average relaxation times $\tau$ could be characterized by power laws, $\tau\sim L^{\alpha}$, where $\alpha$ depends on $p$. Based on the numerical values of $\alpha$, we found that the dependency is also a power law, $\alpha\sim p^{-0.05}$, which implies that the average relaxation times decrease for increasing values of $p$.


\section{Acknowledgments}

The author acknowledge thoughtful remarks by anonymous referees which significantly improved the text. This work was supported by the Brazilian funding agency CNPq.


\begin{thebibliography}{40}

\bibitem{axelrod} 
R. Axelrod, J. Conflict Resolut. 41 (1997) 203-226.

\bibitem{baronchelli}
A. Baronchelli, M. Felice, V. Loreto, E. Caglioti, L. Steels, J. Stat. Mech (2006) P06014.

\bibitem{holley}
R. Holley, T. Liggett, Ann. Probab. 3 (1975) 643-663.

\bibitem{galam}
S. Galam, Int. J. Mod. Phys. C 19 (2008) 409-440.

\bibitem{sznajd}
K. Sznajd-Weron, J. Sznajd, Int. J. Mod. Phys. C 11 (2000) 1157-1165.

\bibitem{loreto_rmp}
C. Castellano, S. Fortunato, V. Loreto, Rev. Mod. Phys. 81 (2009) 591-646.

\bibitem{sznajd_review}
K. Sznajd-Weron, Acta Phys. Pol. B 36 (2005) 2537-2547.

\bibitem{adriano}
D. Stauffer, A. O. Sousa, S. Moss de Oliveira, Int. J. Mod. Phys. C 11 (2000) 1239-1245.

\bibitem{sabatelli}
L. Sabatelli, P. Richmond, Int. J. Mod. Phys. C 14 (2003) 1223-1229.

\bibitem{meu_pla}
N. Crokidakis, F. L. Forgerini, Phys. Lett. A 374 (2010) 3380-3383.

\bibitem{adriano2}
A. O. Sousa, T. Yu-Song, M. Ausloos, Eur. Phys. J. B 66 (2008) 115-124.

\bibitem{auto}
A. A. Moreira, J. S. Andrade and D. Stauffer, Int. J. Mod. Phys. C 12 (2001) 39-42.

\bibitem{raimundo}
R. N. Costa-Filho, M. P. Almeida, J. S. Andrade Jr., J. E. Moreira, Phys. Rev. E 60 (1999) 1067-1068.

\bibitem{barabasi}
R. Albert, A.-L. Barab\'asi, Rev. Mod. Phys. 74 (2002) 47-97.

\bibitem{sznajd_elections1}
A. T. Bernardes, U. M. S. Costa, A. D. Araujo, D. Stauffer, Int. J. Mod. Phys. C 12 (2001) 159-167.

\bibitem{sznajd_elections2}
A. T. Bernardes, D. Stauffer, J. Kert\'esz, Eur. Phys. J. B 25 (2002) 123-127.

\bibitem{sznajd_elections3}
M. C. Gonzalez, A. O. Sousa, H. J. Herrmann, Int. J. Mod. Phys. C 15 (2004) 45-57.

\bibitem{mazzitello1}
K. I. Mazzitello, J. Candia, V. Dossetti, Int. J. Mod. Phys. C 18 (2007) 1475-1482.

\bibitem{mazzitello2}
J. Candia, K. I. Mazzitello, J. Stat. Mech. (2008) P07007.

\bibitem{rodriguez}
A. H. Rodr\'{\i}guez, Y. Moreno, Phys. Rev. E 82 (2010) 016111.

\bibitem{pabjan}
B. Padjan, A. Pekalski, Physica A 387 (2008) 6183-6189.

\bibitem{gonzalez1} J. C. Gonz\'alez-Avella, M. G. Cosenza, K. Klemm, V. M. Egu\'iluz, M. S. Miguel, J. Artif. Soc. Soc. Simul. 10 (2001) 9. Available at http://jasss.soc.surrey.ac.uk/10/3/9.html.

\bibitem{gonzalez2} J. C. Gonz\'alez-Avella, M. G. Cosenza, K. Tucci, Phys. Rev. E 72 (2005) 065102(R).

\bibitem{stauffer_review}
D. Stauffer, J. Artif. Soc. Soc. Simul. 5 (2001) 1. Available at http://jasss.soc.surrey.ac.uk/5/1/4.html

\bibitem{schneider}
J. J. Schneider, Int. J. Mod. Phys. C 15 (2004) 659-674.

\bibitem{stanley_book}
H. E. Stanley, \textit{Introduction to Phase Transitions and Critical Phenomena} (Oxford University Press, Oxford, 1971).

\bibitem{la_lama}
M. S. de La Lama, J. M. L\'opez, H. S. Wio, Europhys. Lett. 72 (2005) 851-857.




\end{thebibliography}
\end{document}